\documentclass[aps,twocolumn,amssymb,epsfig]{revtex4}
\usepackage{subfigure}
\usepackage{graphicx}
\usepackage{epsfig}
\usepackage{amsmath}
\usepackage{color}

\begin{document}
\title{Theory of Cricket: Target Scores and Predictability}
\author{Robin de Regt, Ravinder Kumar}
\affiliation{Applied Mathematics Research Centre, Coventry University, Coventry, UK
}
\begin{abstract}
We propose a model for recalculating the target score in rain affected matches based on empirical data. During the development of the current stage of the Cricket, different methods have been introduced to recalculate the target scores in interpreted games. Currently, the International Cricket Council (ICC) uses the Duckworth-Lewis method and have in the past strongly considered changing to the VJD method. Here, 
we introduce a simple approach to calculate target scores in interrupted games by considering the area under a run rate curve. To calculate the target we have analysed over a decades worth of empirical data using various statistical methods.
As in the case of Duckworth-Lewis method, we also have two parameters in our model, that is overs and wickets in combination. We also found that in the one day international cricket (ODI) wickets play a crucial role whereas in T20 cricket they do not effect the run rate of the games to the same degree. Using empirical and mathematical arguments we show that the run scoring distributions are independent of the innings.
\end{abstract}
\maketitle

\section{Introduction}
\vspace{-0.2cm}
In the game of cricket determining how best to re-evaluate a score when a game has been interrupted by inclement weather has been of much debate for many years. There have been a number different methods proposed to decide the target score for interrupted games \cite{clarke88,rego95,VJD,ducklewis}. The incumbent method used by the governing body of world cricket, the ICC, is the Duckworth Lewis method. However it has come under much criticism especially in recent times for it ability to recalculate scores mainly in T$20$ cricket where the number of fallen wicket do not seem to play as an important role as it does in the longer format of ODI cricket \cite{DLfail}.   
In this introduction we briefly discuss some of the different models used to recalculate the target score in the event of interrupted games where overs are lost. Some of these methods are discussed in \cite{ducklewis} in more detail. Most of the methods consider basic algorithms to recalculate target scores by implement simple arithmetic, whereas other methods like Parabola, VJD, Clark curves and Duckworth and Lewis are slightly more involved. All have their advantages as simple methods allow for easy understand for laymen whereas more detailed and sophisticated methods are more likely to readjust targets scores in a fairer manner. Here we will discuss the two main competing methods the ICC have considered using for cricket; the Duckworth Lewis method and VJD method. 

\subsection{Duckworth-Lewis (DL) Method}
This DL method uses the idea that teams have two main resources from which to score their runs, these being the number wickets and overs remaining to them. At the start of an innings a team has all of it resources available to them and these are reduced as the overs are bowled and wickets are lost. The equation that DL have used to describe this relationship is given in equation $(1)$. 
\begin{equation}\label{1}
Z(u)=Z_0(w)[1-\exp(-b(w)u)]
\end{equation} 
where $Z_0(w)$ is the asymptotic average score and $b(w)$ is the exponential decay for $w$ fallen wickets in $u$ given overs. The parameters for this function 
are not published for commercial reasons however they do stipulate that parameters used have come from extensive research on many ODI games and can attest that their model produces reasonably results under practical circumstances. For purposes of illustration we have created fig $(1a)$ to demonstrate how these equation might look qualitatively.

To determine the proportion of resources available one can simply calculate the ratio between maximum available resources and resources at a particular point in time of an innings. From these proportions tables can be produced to assist in the readjusting of scores in reduced games. For a more detail description of the model we refer you to ref \cite{ducklewis}.      
\begin{figure}[!htb]\label{fig1}
\centering
 \subfigure[ ~Duckworth]{\includegraphics[scale=0.5]{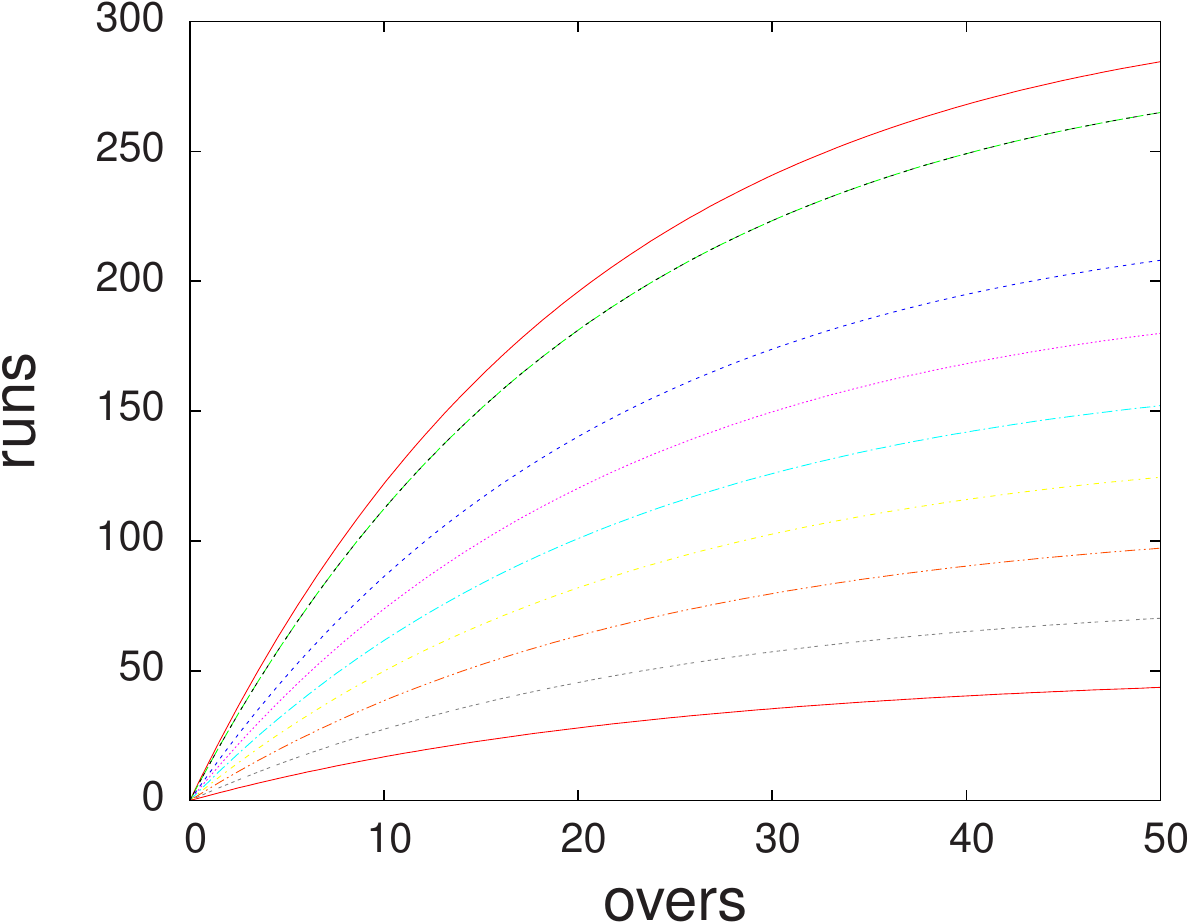}}\\
 \subfigure[ ~VJD]{\includegraphics[scale=0.5]{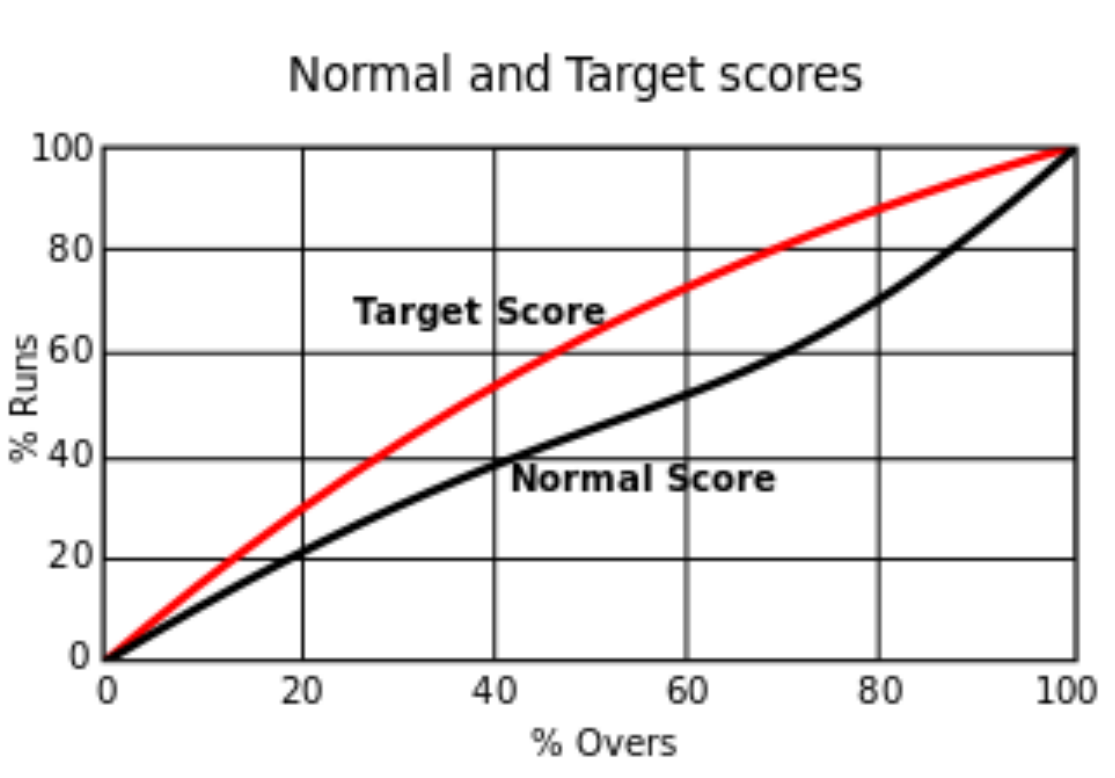}}
\caption{Using our own parameters for decay exponential and average runs scored we plot DL methods}
\end{figure}

 \begin{figure*}[t]
        \centering
        \subfigure[ ~IPL 1st inning]{\includegraphics[scale=0.4]{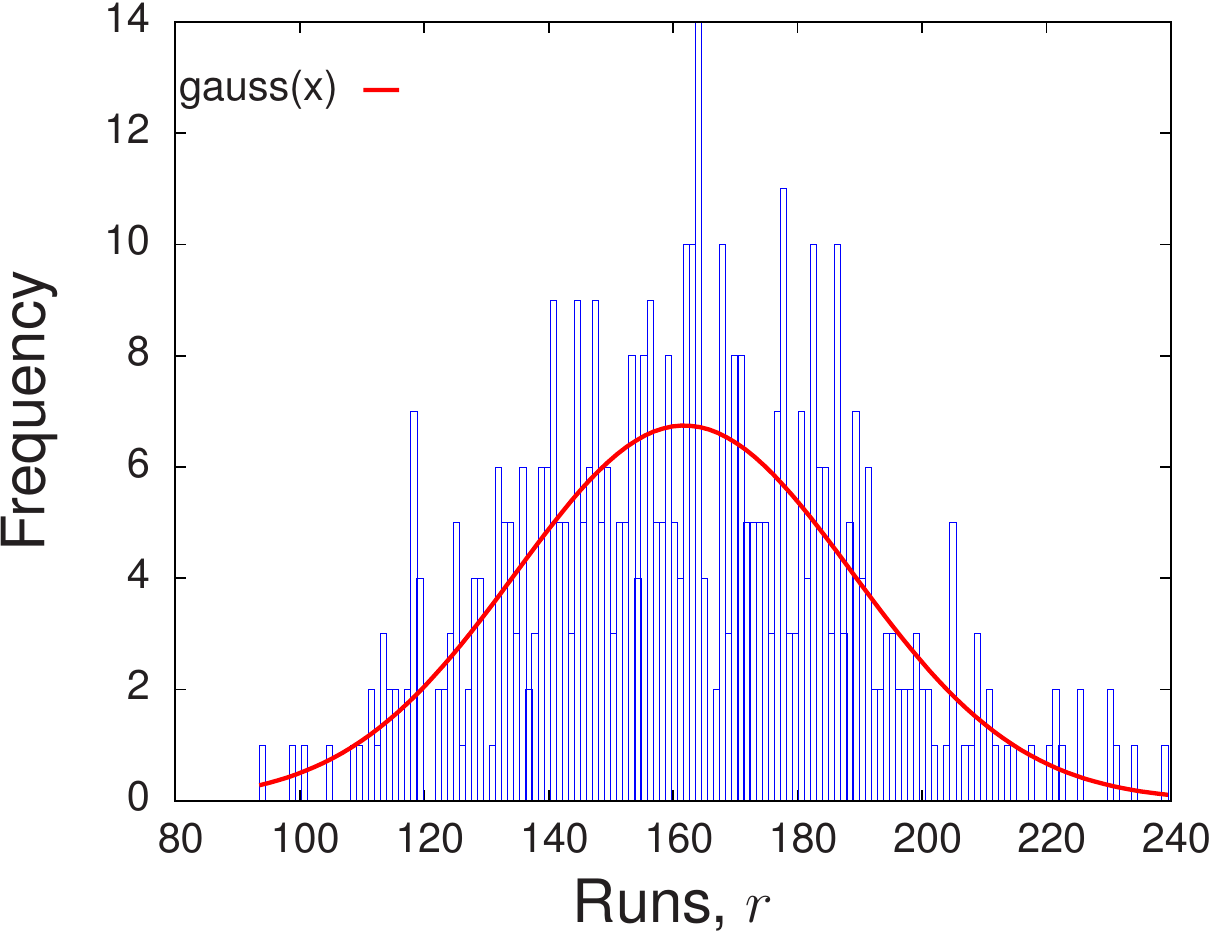}\label{T0_step}}
        \subfigure[ ~T20I 1st inning]{\includegraphics[scale=0.4]{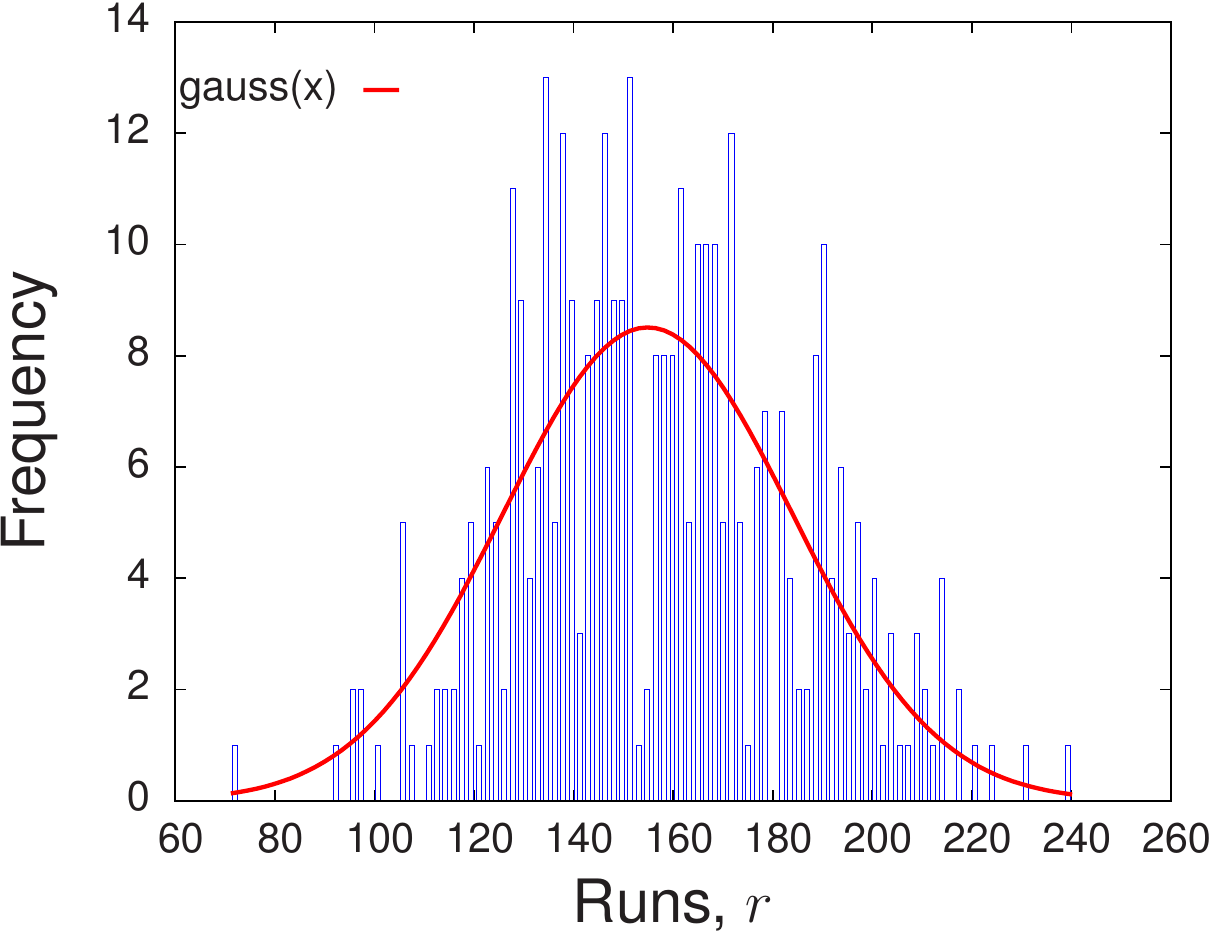}\label{T0_sin}}
        \subfigure[ ~ODI 1st inning]{\includegraphics[scale=0.4]{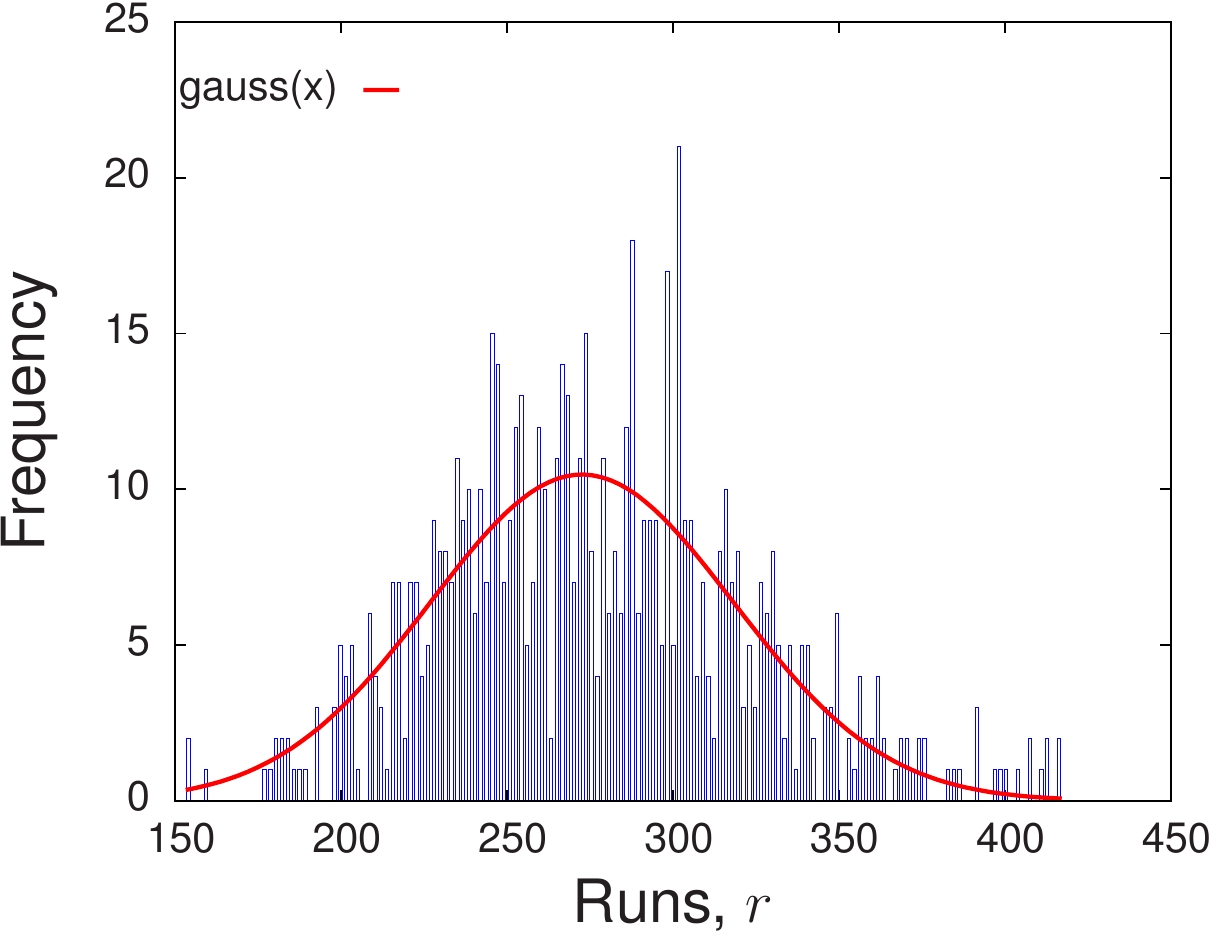}\label{T0_odl}}
         \subfigure[ ~IPL 2nd inning]{\includegraphics[scale=0.4]{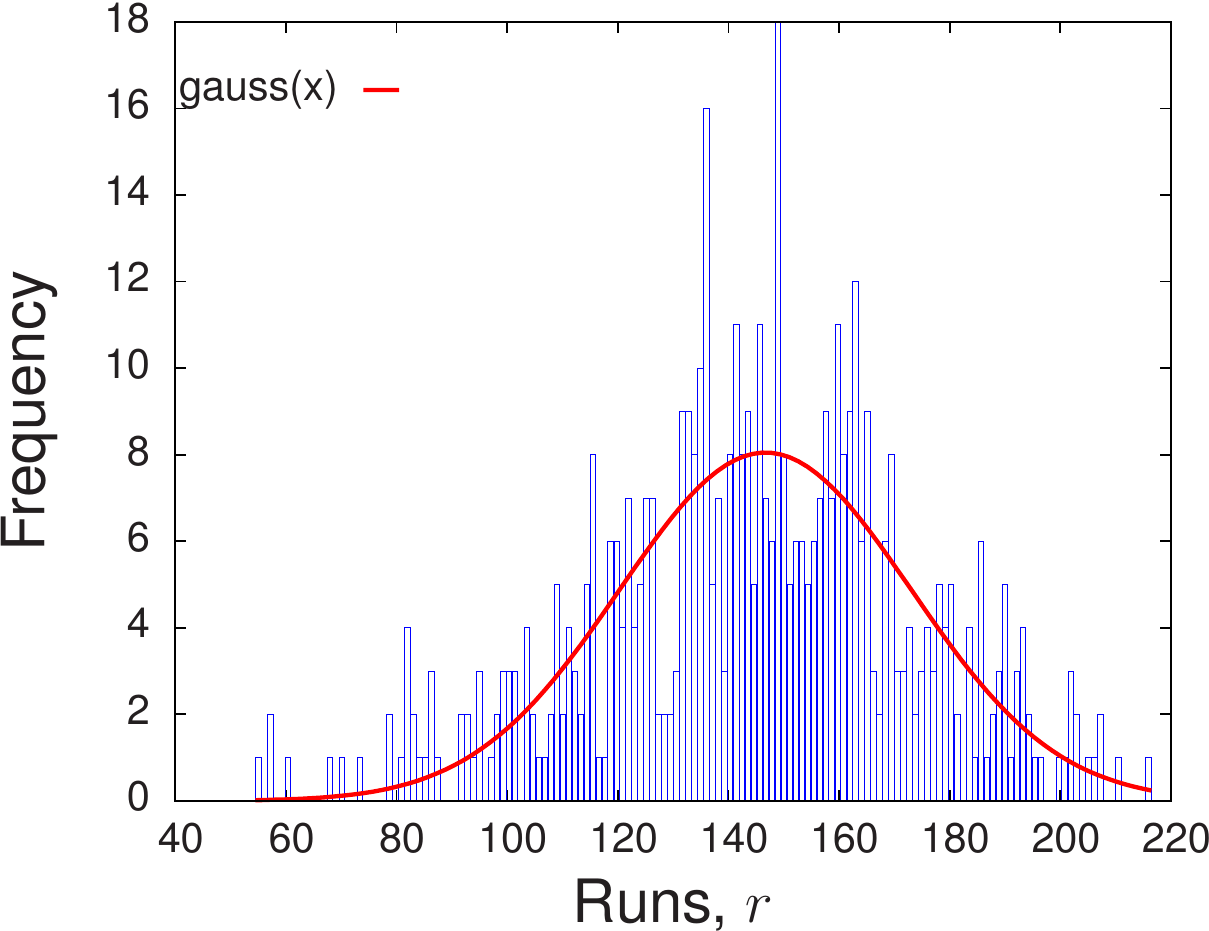}\label{T1_step}}
        \subfigure[ ~T20I 2nd inning]{\includegraphics[scale=0.4]{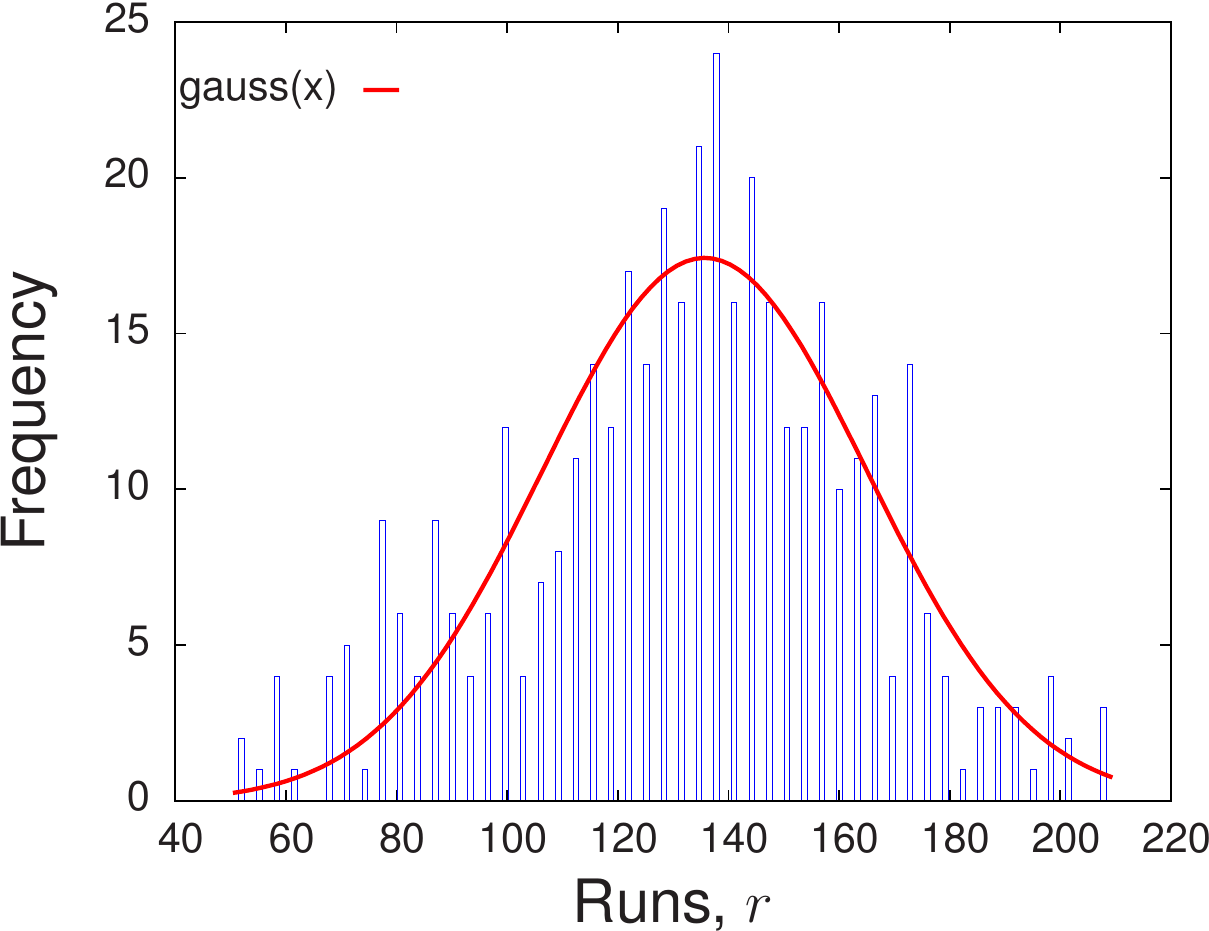}\label{T1_sin}}
        \subfigure[ ~ODI 2nd inning]{\includegraphics[scale=0.4]{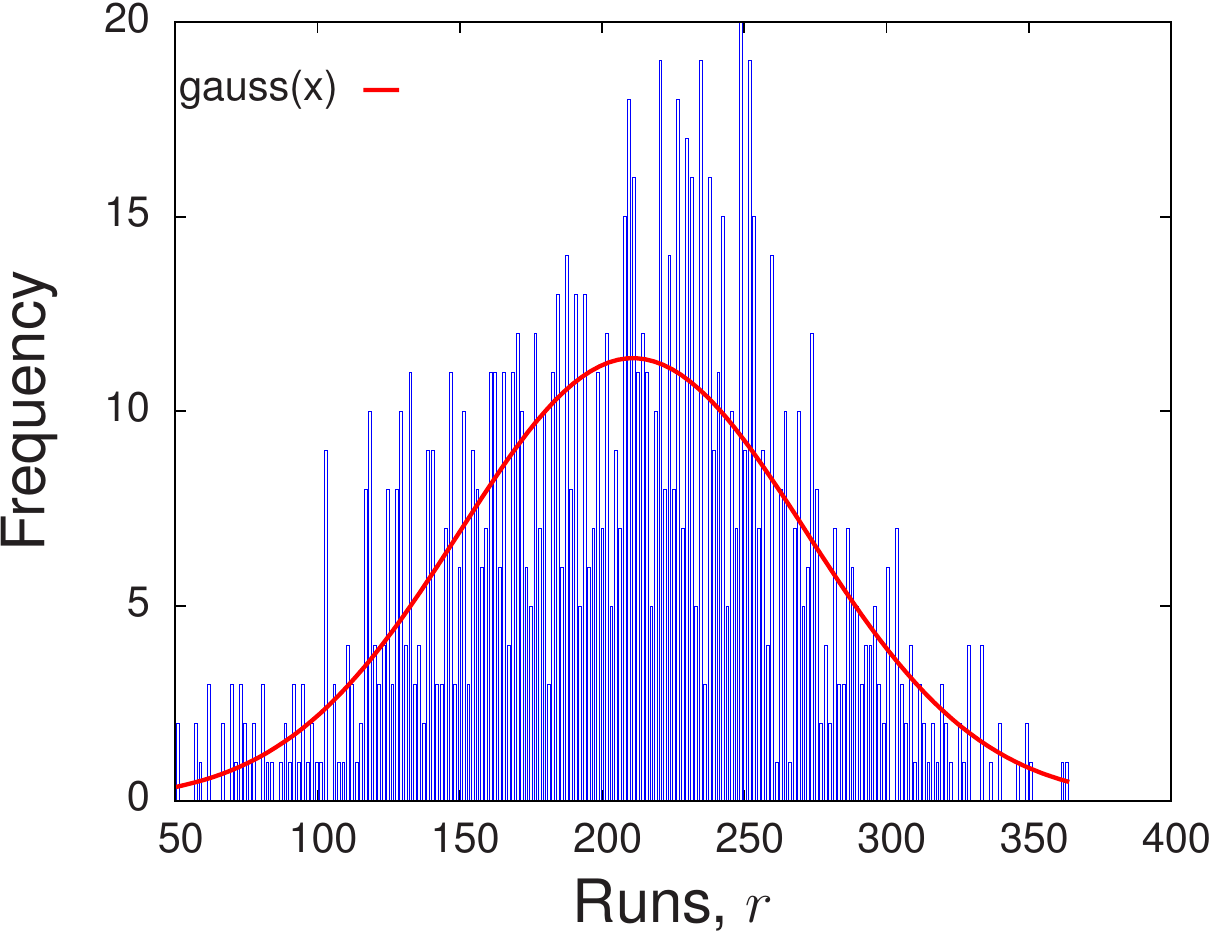}\label{T1_odl}}
        \caption{Comparative histograms of the total run scored for the different formats of the cricket. These plots show that at the average scoring rates are distributed normally irrespective of the inning or format of the cricket.}
\label{fig:hist_all}
\end{figure*}
\subsection{VJD Method}
The VJD method is based on using two different types of curves. One curve, the normal curve, is for the team batting first and considers how a teams batting first tend to score their runs knowing that they have their full allocated overs to complete and hence can plan their innings more appropriately. To determine the normal curve the VJD method breaks the innings down into seven phases: $0-5$,$6-15$,$16-25$,$26-30$,$31-40$,$41-45$ and $46-50$. Then by using a regression equation the cumulative curve is determined. The second curve, called the target curve, is found by considering the first innings breakdowns. Which is done by finding and reordering these first innings. This can be visualised in fig $1b$ and for a more in depth breakdown of this method the reader is referred to \cite{VJD}.     
 \begin{figure*}[tb]
        \centering
       \subfigure[ ~IPL 1st inning]{\includegraphics[scale=0.4]{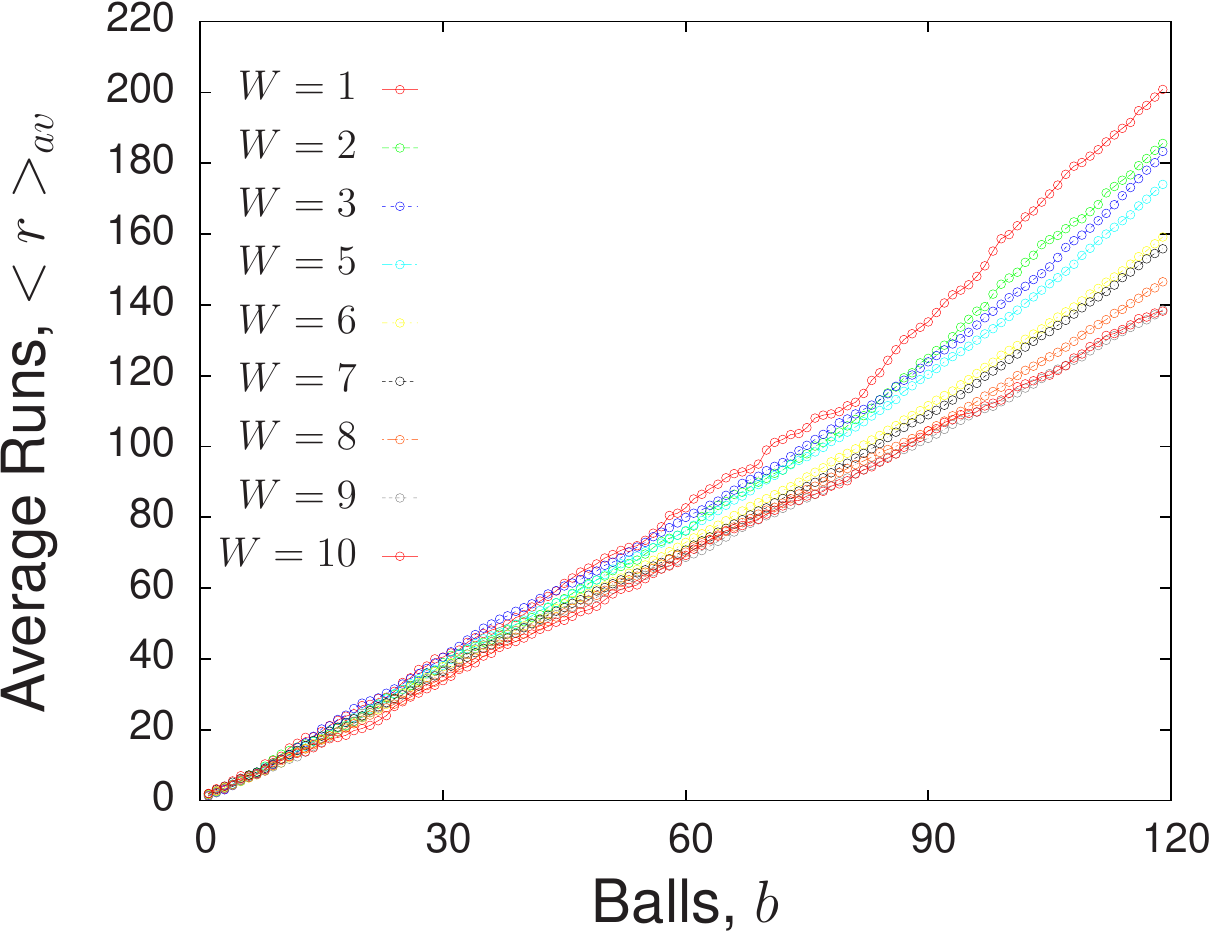}\label{T0_step}}
        \subfigure[ ~T20I 1st inning]{\includegraphics[scale=0.4]{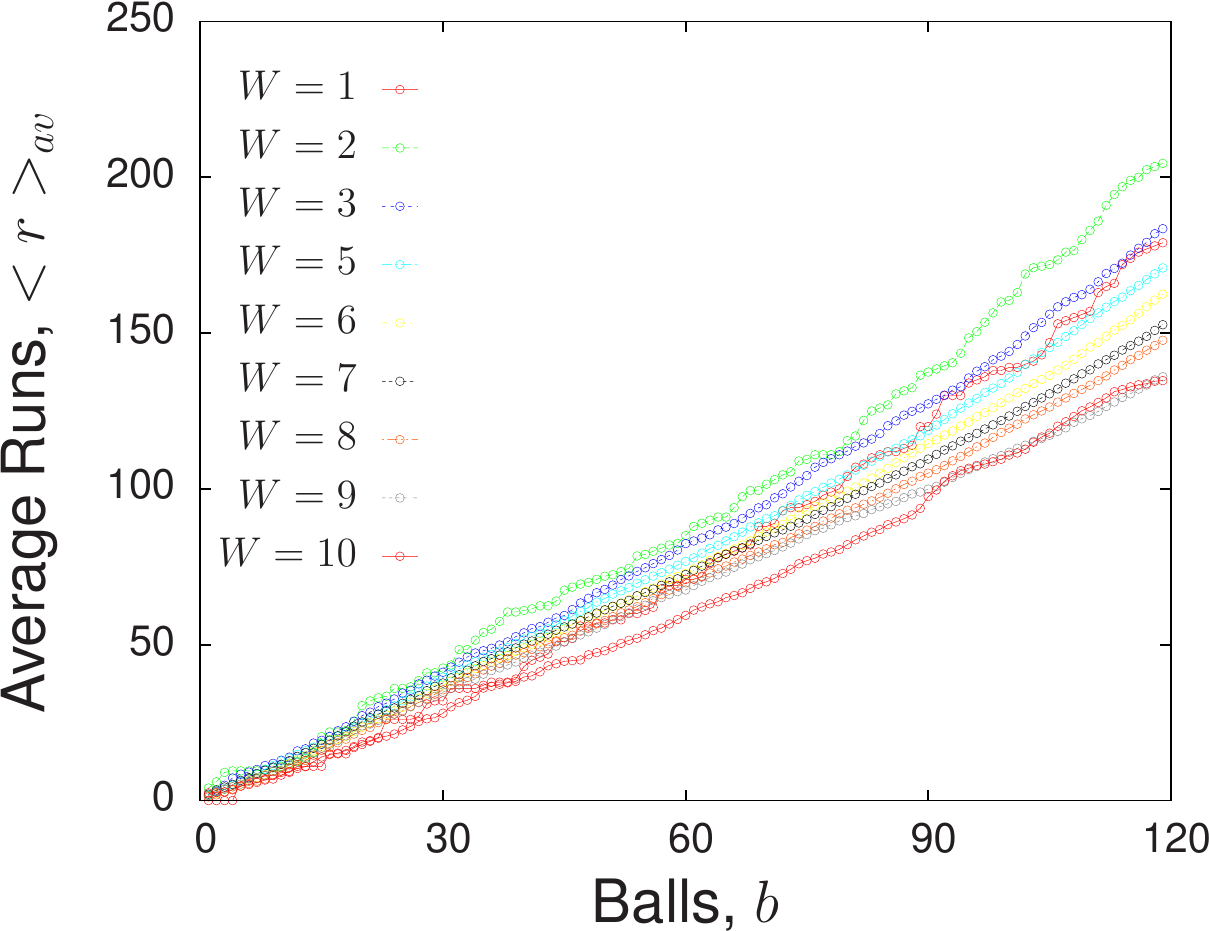}\label{T0_sin}}
        \subfigure[ ~ODI 1st inning]{\includegraphics[scale=0.4]{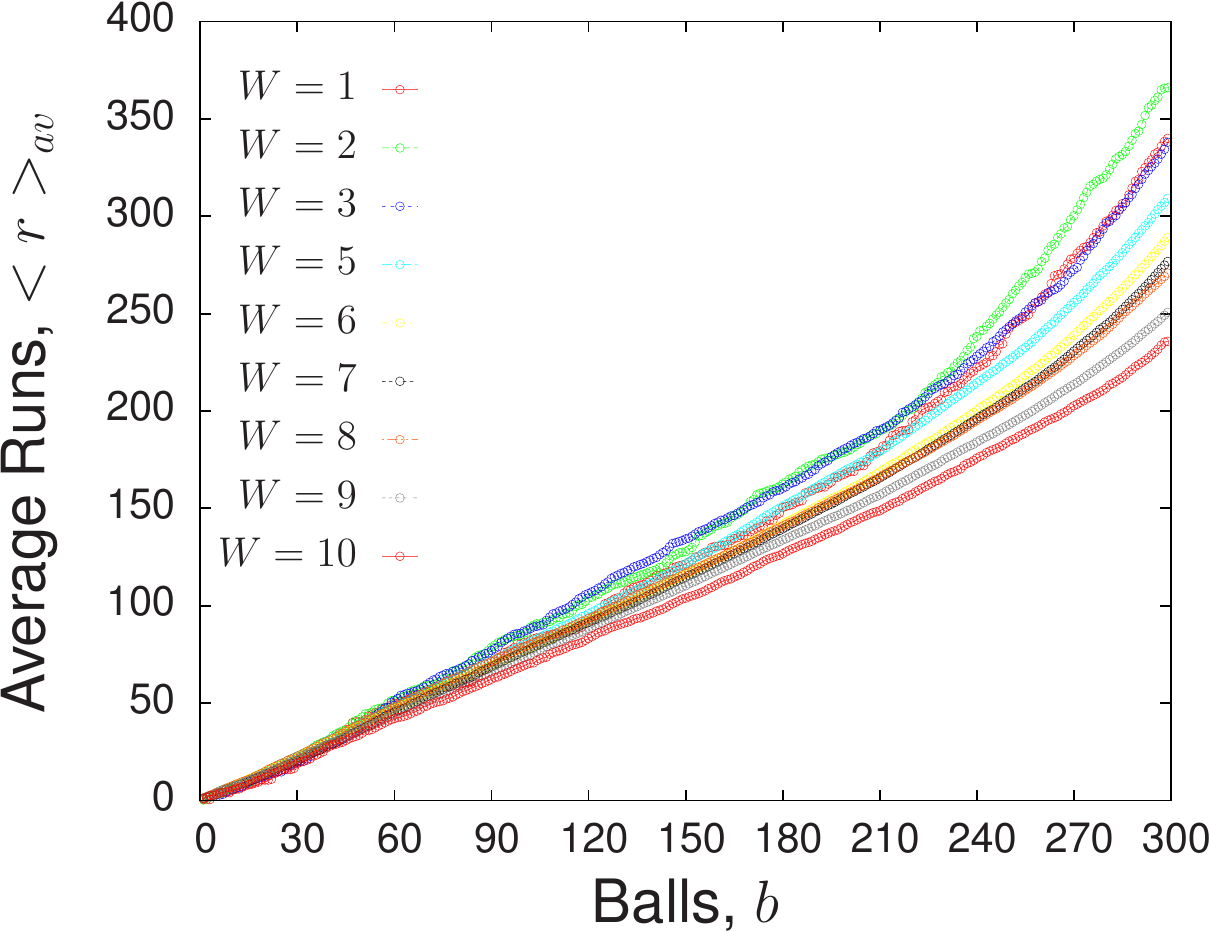}\label{T0_odl}}
        \subfigure[ ~IPL 2nd inning]{\includegraphics[scale=0.4]{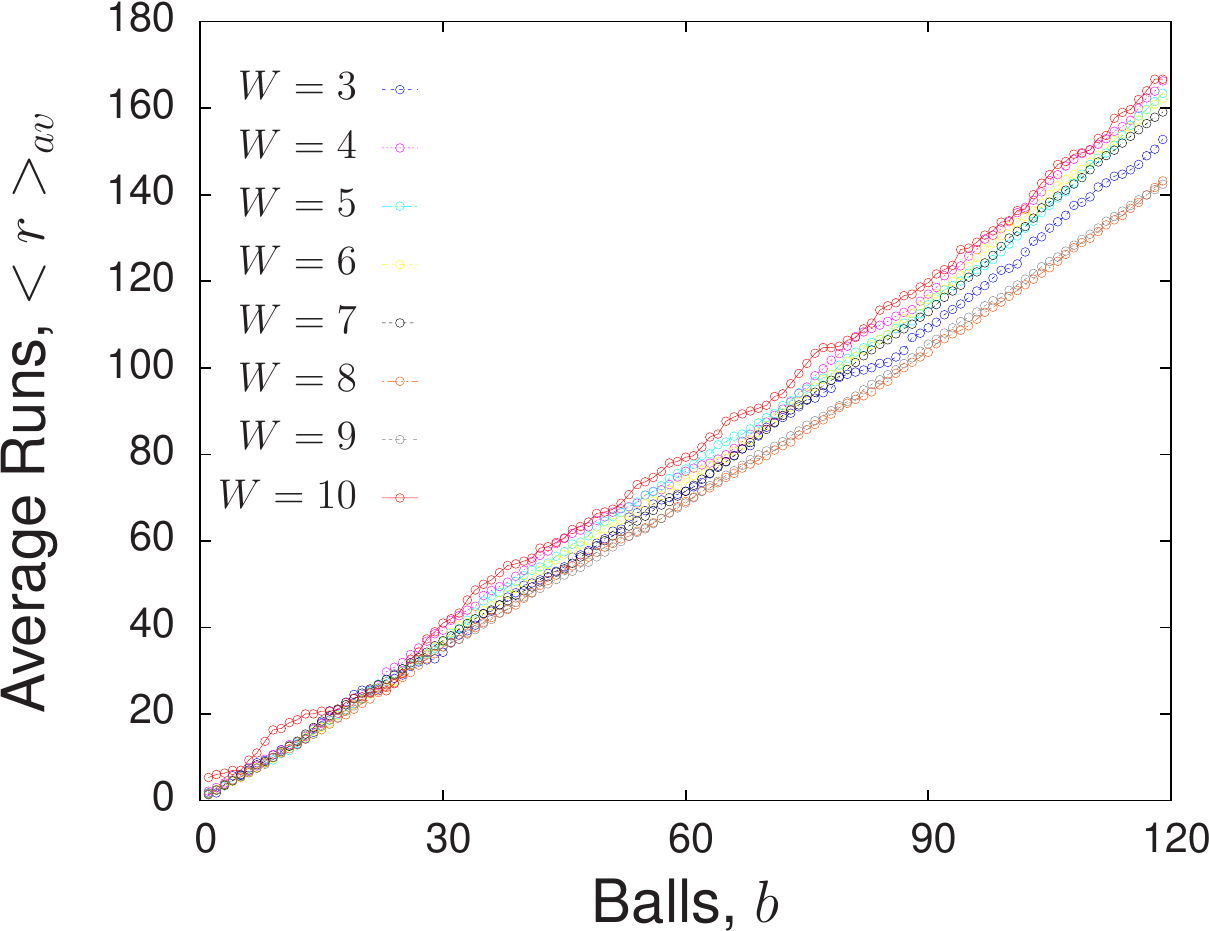}\label{T1_step}}
        \subfigure[ ~T20I 2nd inning]{\includegraphics[scale=0.4]{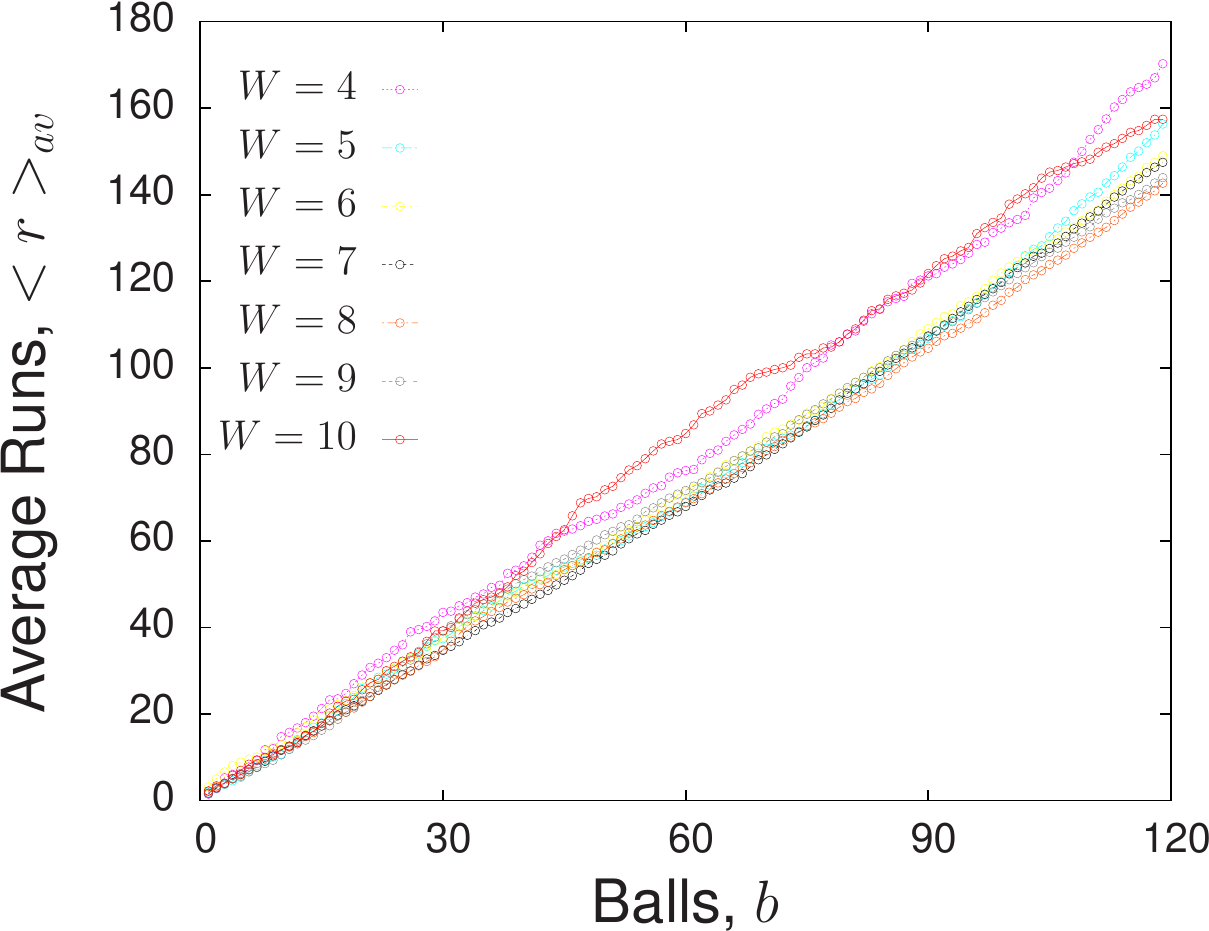}\label{T1_sin}}
        \subfigure[ ~ODI 2nd inning]{\includegraphics[scale=0.4]{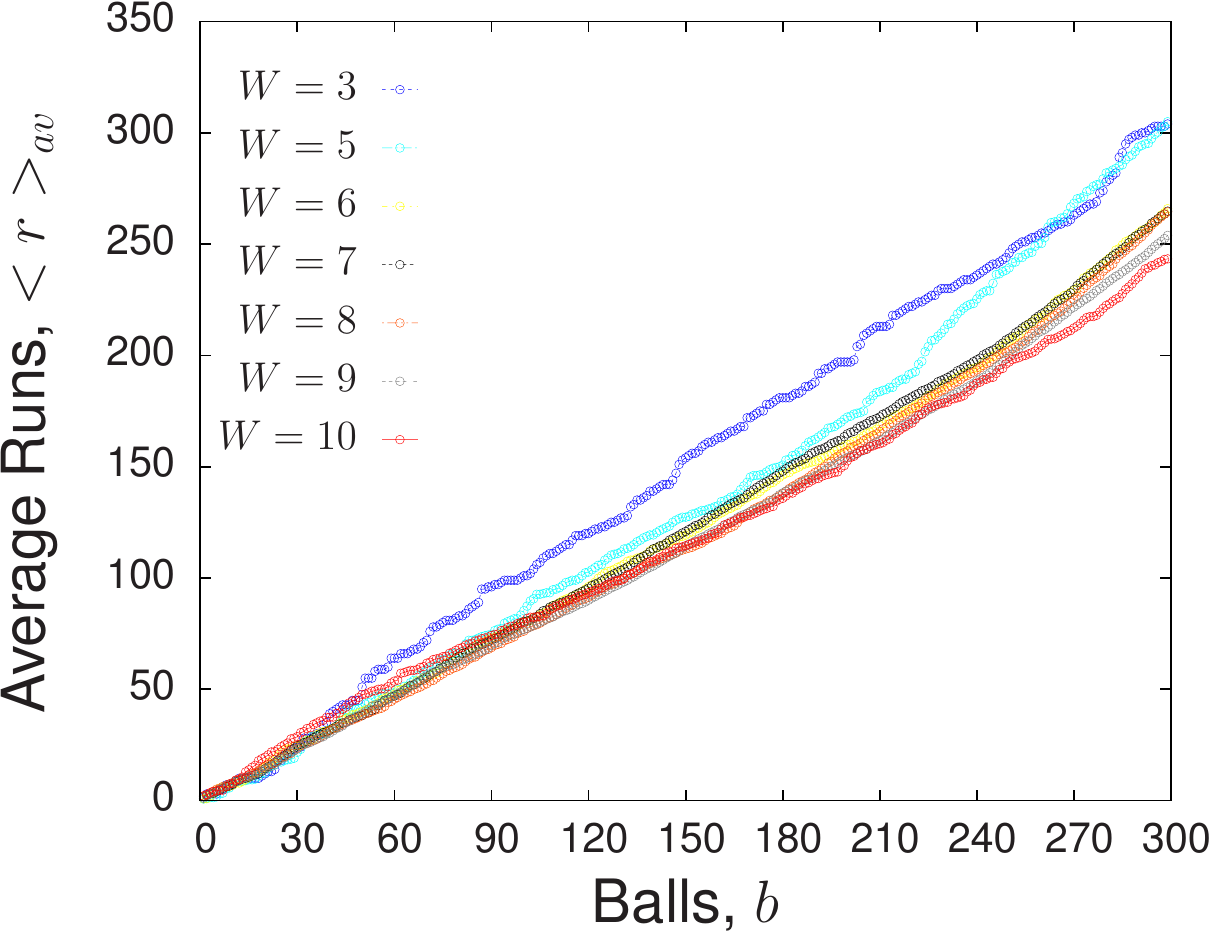}\label{T1_odl}}
        \caption{Comparative histograms of the total run scored for the different formats of the cricket. These plots show that at the average scoring rates are distributed normally irrespective of the inning or format of the cricket.}
\label{fig:av_all}
\end{figure*}

\section*{Data Analysis \& Results}

The data for this analysis has come from the Cricsheets website which can be found at \cite{cricsheets}. The matches it covers among others include $1161$ ODI, $454$ T$20$I and $517$ IPL matches. The database has a large number of games from as far back as $2005$ right up until recent matches. The information that is provided is on a ball by ball basis where they document who bowled,batted, was the non striker and the event of every ball i.e runs score or wicket taken. It also provides general information about where the games were played, against whom and the umpires adjudicating the game. 

With this type of information one could use this to analyse scoring rates empirically for the various formats of cricket considering a number of factors like for example the ground or country of match or how number of fallen wickets affect the scoring rates. To analyse the scoring rates we have only considered the first innings of matches for simplicity and may at a later stage consider the second innings. The reason being in most cases teams batting first can or at least have the opportunity to complete their entire allotment of overs and hence run rates for the whole innings can be acquired. Whereas in the case of the team batting second if they win then they will almost surely do this before completing their overs which might affect run rate analysed toward the end of second innings. In order to justify our method we look at the histograms of scored runs from the empirical data. The total scored runs follow a normally distributed behavior. Mathematically standard distribution is given by:
\begin{equation}\label{dist-norm}
\frac{A}{\sqrt{2\pi}\sigma}e^{-\frac{(x-\xi)^2}{2\sigma^2}},
\end{equation}
where $\sigma$ is the standard deviation in the data set, $\xi$ is the mean of the data set interpreted as the average score of all the games and $A$ is the amplitude. The histograms of the runs for IPL, T20I and ODI format of the game are given in Fig$(2)$. The data is then fitted using Eq(\ref{dist-norm}). The total runs scored are distributed normally irrespective of the innings or game format, but the average score in the second innings is less when compared to that of the first innings. This fact can be understood by assuming that probability of the team playing first or second innings is 50\%, but the teams batting in the second innings rarely complete alloted overs. If the team batting second looses, then either they were bowled out before completing their allotted overs or their run rate was less then the team batting first. If 50\% of games are lost by team batting second this would reduce the total average score significantly because in the cases that they win that would only have to pass the first teams score and hence would not continue any further as they have won the game. Hence, the fact that the average score in the second innings is less than that of the first is merely a statistical fact and does not mean that the second team always lose. The values of $\sigma,\xi$ and $a$ are listed in the table($1$).
\begin{table}
\begin{tabular}{|l|c|c|c|r|}
  \hline
  Format & Inning &$\xi$ &$\sigma$&$A$\\
  \hline
   IPL & Ist & 161.785  & 27.0883 & 458.055 \\ \cline{2-5}
       & 2nd & 146.692  &26.3473 &531.573 \\
  \hline 
   \hline
   T20 I & Ist &154.887 & 29.0680 &620.049  \\ \cline{2-5}
    & 2nd  & 136.557  & 29.4560  &642.740 \\
  \hline 
    ODI & Ist& 272.538  & 45.8380  &1202.920 \\ \cline{2-5}
    & 2nd  &  210.893 & 61.0630  &1739.750 \\
  \hline
   \end{tabular}
  \caption{ Tabular values for the average score $\xi$, standard deviation $\sigma$ and amplitude $a$.}
\end{table}

We can see from Figs($3$a-f) that for $20$ over matches the run rates are nearer to constant then that of $50$ matches where in these matches run rates tend to increase through out an innings; especially after the $210$ ball mark.  





\section*{Proposed Method for recalculating scores}
We calculated average run rates for each innings of considered formats of the cricket for certain number of wickets fallen using available empirical data. A plot corresponding to number of wicket fallen is shown in Fig(\ref{fig:av_all}). The data was fitted by using a polynomial function of the form:
\begin{equation}\label{fit:eq}
 f(x)=ax^3+bx^2+cx+d,
\end{equation}
where variable $x$ describes the number of balls and $a,b,c$ and $d$ are constants. We know that at $x=0$ balls the score should be zero. This fact is used to set $d=0$. Although cubic fit works for all the data but the data for T20 international and IPL can be fitted using a quadratic function as well. From Fig.~\ref{fig:av_all}, one observes that the scoring rates are very high closer to the end of the innings. 
\begin{figure}[t]
\includegraphics[scale=0.5]{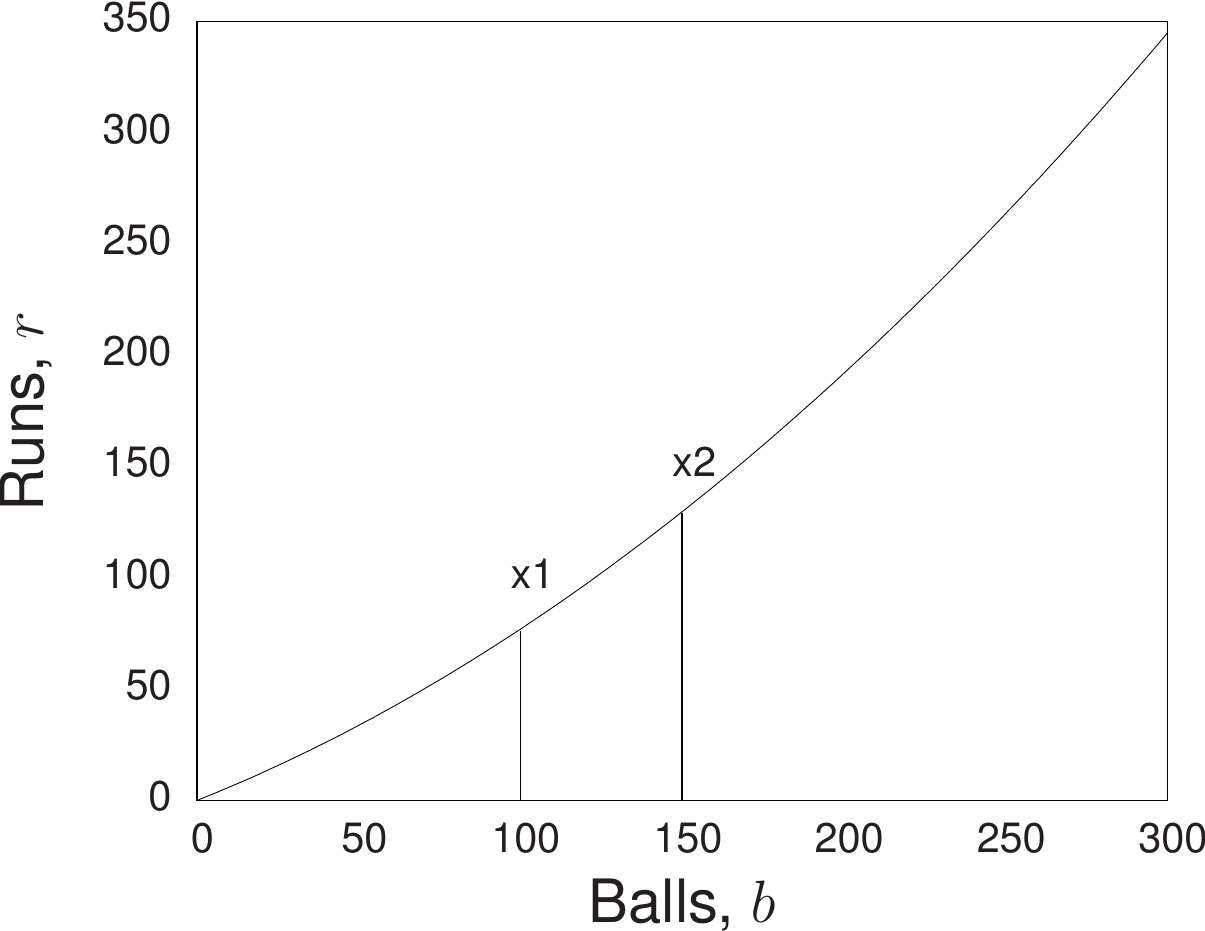}\label{fig5}
\caption{Graphical illustration of to recalculate scores}
\end{figure}
The DL and VJD methods mentioned earlier use exponential and regression fits to reset the target score of an interrupted game respectively. Here we propose a new method where we use area under the curve of fitted run rates to reset the target. We use Eq.(\ref{fit:eq}) to calculate area under the curve. It seems that this method violates assumption mentioned in \cite{ducklewis} that the method to reset the target should be simple and should not require more than a pocket calculator for the calculations. This assumption is not valid any more because of the use of computers nowadays, but nevertheless we will show that the model proposed here needs nothing more than a pocket calculator to reset the target. This is explained in a working example later. The area under a curve can be calculated by integrating the function given in Eq(\ref{fit:eq}). The area under a curve for a non interrupted game with $N$ balls is given by:
\begin{equation}
 A_{\rm game}=\int_{0}^N (ax^3+bx^2+cx)dx= \dfrac{N^2\left(N\left(3aN+4b\right)+6c\right)}{12}.
\end{equation}
If a game is interrupted after $n$ balls and restarted again after $m$ balls, the area of remaining curve can be recalculated by:
\begin{equation}
A_{\rm int}=\int_{0}^n (ax^3+bx^2+cx)dx+\int_{m}^N (ax^3+bx^2+cx)dx
\end{equation}
\begin{equation}
\small
A_{\rm int}=\dfrac{3a(n^4+N^4-m^4)+4b(n^3+N^3-m^3)+6c(n^2+N^2-m^2)}{12}.
\end{equation}
 $A_{\rm game}$ and $A_{\rm int}$ are the areas for a non-interpreted and interrupted game respectively. The ratio of these areas is given by:
 \begin{equation}
  R_{\rm target}=\frac{ A_{\rm int}}{ A_{\rm game}}.
 \end{equation}

The ratio $R_{\rm target}$ is used to reset the target ($T_{new}$) for the interrupted games and uses the target score ($T_s$) and current score ($C_s$) as variables to reset the targets.
\begin{equation}
 T_{new}=R_{\rm target}\left(T_s-C_s \right)
\end{equation}
At this point we can mention that this ratio can be calculated by using a pocket calculator. Hence, the assumption of D/L method is fulfilled here. For example in an ODI game if a team batting first scores $275$ and in the second innings the game is interrupted at $20$ over mark with their score on $100$ for the loss of $2$ wickets. Due to the interruption 10 overs are lost leaving only 20 over left in the game. Using the fitted function where only two wickets had fallen, $a$, $b$ and $c$ are $-0.0031$, $1.0298$ and $0$ respectively. We can use equations($4$-$8$) to calculate a revised target score of $230$ runs in the remaining $20$ overs.


\section*{Conclusion and discussion}
We have analysed the data for the Indian Premier League (IPL), T20 international and ODI cricket.
The average scores of the games were found to be distributed normally and independent of the innings or the format of the game. We used ball by ball averages to fit the data and thus allow us to implement some simple integrals to revise target scores. This assisted in the development of a new method to reset the target score of interrupted games. We feel that this simple model could further be improved with more sophisticated data averaging techniques such as moving averages for example which will be the subject for future work. 
\section*{Acknowledgement}
We thank Stephen Rushe of Cricsheet and ESPN Cricinfo for providing the data without which we would not have completed this work. We would further like to thank Martin Weigel for his insightful comments. The financial assistance from Coventry University, Coventry, UK.


\end{document}